\documentclass[a4paper,fleqn]{cas-dc}

\usepackage[numbers]{natbib}
\usepackage[justification=centering]{caption}
\usepackage{enumitem}
\usepackage{etoolbox}
\usepackage{csquotes}
\usepackage{booktabs}
\usepackage{multirow}
\usepackage{lscape}
\usepackage{longtable}
\usepackage{colortbl}
\usepackage{svg}
\usepackage{pdfcomment}
\usepackage{vcell}
\usepackage[table,xcdraw]{xcolor}
\usepackage[dvipsnames]{xcolor}
\usepackage{ragged2e}
\usepackage{array}

\usepackage{enumitem}
\usepackage{etoolbox}
\usepackage{csquotes}
\usepackage{booktabs}
\usepackage{multirow}
\usepackage{lscape}
\usepackage{longtable}
\usepackage{colortbl}
\usepackage{svg}
\usepackage{pdfcomment}
\usepackage{vcell}
\usepackage[table,xcdraw]{xcolor}
\usepackage[dvipsnames]{xcolor}

\def\tsc#1{\csdef{#1}{\textsc{\lowercase{#1}}\xspace}}
\tsc{WGM}
\tsc{QE}

\begin{document}
\let\WriteBookmarks\relax
\def\floatpagepagefraction{1}
\def\textpagefraction{.001}

\shorttitle{}    

\shortauthors{}  

\title [mode = title]{The DataSquad Experiment: Lessons for Preparing Data and Computer Scientists for Work}


\author[1]{Paula Lackie}[orcid=0000-0002-0819-2858]

\cormark[1]
\fnmark[1]

\ead[1]{plackie@carleton.edu}



\affiliation[1]{organization={Carleton College},
            city={Northfield},
            postcode={55057}, 
            state={MN},
            country={USA}}


\author[2]{Elliot Pickens}[orcid=0009-0008-5705-7207]

\ead[2]{epickens@seas.upenn.edu}

\affiliation[2]{organization={University of Pennsylvania},
            city={Philadelphia},
            postcode={19104}, 
            state={PA},
            country={USA}}


\author[1]{Dashiell Coyier}[orcid=0009-0003-1650-9370]

\ead[3]{coyierd@carleton.edu}

\begin{abstract}
The DataSquad at Carleton College addresses a common problem at small liberal arts colleges: limited capacity for data services and few opportunities for students to gain practical experience with data and software development. Academic Technologist Paula Lackie designed the program as a work-study position that trains undergraduates through structured peer mentorship and real client projects. Students tackle data problems of increasing complexity—from basic data analysis to software development—while learning FAIR data principles and open science practices. The model's core components (peer mentorship structure, project-based learning, and communication training) make it adaptable to other institutions. UCLA and other colleges have adopted the model using openly shared materials through "DataSquad International." This paper describes the program's implementation at Carleton College and examines how structured peer mentorship can simultaneously improve institutional data services and provide students with professional skills and confidence.
\end{abstract}



\begin{keywords}
 \sep Experiential learning \sep Research software engineering \sep Data services \sep Student workers 
\end{keywords}

\maketitle

\section{Introduction}
Small institutions like Carleton College struggle to support remarkably diverse and growing data and programming needs.\cite{kellam_databrarianship_2016} While US academic libraries now lead in the general support of data and research support services on campuses\footnote{e.g.\href{https://www.library.ucla.edu/visit/locations/data-science-center}{ UCLA Library Data Science Center}, \href{https://library.princeton.edu/services/data-and-statistical-services}{Data and Statistical Services - Princeton University Library}, \href{https://library.duke.edu/data}{Duke's Center for Data and Visualization Sciences}, etc.}  longer/larger projects, such as research software development, remain underdeveloped.\cite{verma_research_2023}  Additionally, higher education administrative offices are typically unable to take advantage of the services designed to support academics and their research.  (e.g. providing programmatic support to facilitate the analysis of IT-helpdesk ticket patterns, de-tangling Google metadata from Shared Drives to help determine the viability of remaining content, creating a sustainable and visual database infrastructure for Theater Dept props\footnote{These are all examples of DataSquad projects at Carleton.}) 

As Research Software Engineering (RSE) emerges as a discipline in its own right, knowledge about training pathways remains limited, with early initiatives often relying on anecdotal evidence rather than assessment.\cite{Lamprecht2022} A student employment process with a skilled advisor may help bridge the gap between capable—but stretched—staff and the service and support needs of a campus. However, working with a steady stream of novices has its own set of obstacles.  These challenges extend to research software engineering, where practitioners need both technical depth and collaborative skills like documentation and communication.\cite{Cohen_2021_four_pill, Goth_2024_found_comp}

In 2020, \href{https://iassistdata.org}{IASSIST} members surveyed data support professionals globally to discern whether there were shared obstacles in using students to help support data services.\cite{wiltshire_using_2025} The survey data were analyzed by William Foote (UCLA '22), a member of UCLA's DataSquad.\cite{Foote2022-blog}  From Foote's summary, we learn that the primary perceived obstacles to using students to supply data services include: they are too short-term, staff do not have time to train them, and once trained, the best get hired elsewhere. The effect is a sense of constant churn in the delivery of  these services.

The IASSIST collaboration on working with students was the start of DataSquad International: an aspirational network of data support managers who wish to employ students in the delivery of their services. To help mitigate the obstacles to managing students, one goal was for the network to freely distribute the flexible DataSquad model and its openly shared infrastructure; a handbook, website design, job descriptions and advertising materials, logo/s, and etc.  The pandemic of 2020-2023 put these plans on hold while data-support needs persisted.

From another angle, students face the initial obstacle of needing to have experience to get an internship to get experience to then get a job.  The entry points are few and highly competitive, adding an extra burden to students from disadvantaged backgrounds and further reinforcing  the lack of diversity among people employed in data and tech careers.\cite{GeorgetownCEW2024}

Compounding the challenge of how to get early practical experience is the widely recognized data skills gap between academic training and workforce demands.\cite{wiltshire_using_2025} Time constraints and the breadth of material typically force faculty to focus on teaching specific technical skills.  As important as it is, there simply is not space in the curricula to adequately prepare students for the nuanced and complex realities of real-world data, software preparation, management, and analysis. Changes to curricula and an emphasis on experiential learning can help, but there is no silver bullet.\cite{Kang2022} This gap results in a steep learning curve for new employees and researchers entering technical (data or software engineering) roles. 

The original DataSquad International members recognized the potential in sharing the perceived burden of helping iterations of fresh novices get supported experience to be competitive in the internship world.  They also need to deliver a service to those who seek data (or programming) support.  With planning, the two needs can be complementary. To make this pairing work, a critical transformation in perception is necessary: It is essential that employers recognize that students themselves are a part of the service provided.  Skilled students can disseminate best practices (following FAIR principles and Open Science methods) across campus and beyond.\footnote{FAIR principles in research software have seen some recent development.\cite{fair-rse}} While the more direct services (e.g. data cleaning, wrangling, and visualization) may be slower in delivery than with a dedicated professional staff, few institutions have this luxury.  With students, more clients can have the opportunity to have their problems and projects addressed. 

We would be remiss to ignore current discussions on LLM-based AI tool use\footnote{ Here we are deliberately using a catch-all term as specifying a model is unimportant to this discussion.} in schools.  The dialogue ranges from claims of AI as Armageddon to Pollyanna.\footnote{The dialogue is so rapid as to make citation superfluous.} Experience will likely reveal that it is all possible; only the scale of negative to positive impact will vary.  The usual supposition remains that students who heavily rely on AI tools without also engaging with their own critical skill development will miss foundational opportunities to develop human-based skills necessary for their own success (as well as for our collective future!).  Anecdotally, for this article, first-year, first-term co-author Dashiell Coyier used AI coding support to learn R, to implement his ideas for data visualizations, enhance his capabilities, and compound his knowledge in less than 8 weeks.  He has been self-regulating in applying his critical thinking to all of his AI interactions and his results have been very rewarding. 
\section{The DataSquad Model}

The DataSquad has been developed by Paula Lackie over 30 years of supporting \href{https://www.carleton.edu/about/}{Carleton College} with its data needs. At the beginning of her tenure at Carleton, Lackie handled a majority of research data support on her own. But as demand increased on campus, Lackie hired student workers to help, soon recognizing that guiding eager students was more beneficial than working on individual support projects herself.  Through trial and error with Carleton students and their projects, the DataSquad emerged.  By 2018, the basic roles and responsibilities were solidified into the experimental DataSquad Model.\cite{wiltshire_meeting_2025} 

The employment of students on the \href{https://carletondatasquad.bitbucket.io}{Carleton DataSquad} is a win-win-win situation. By leading students through project-support, the DataSquad can help significantly more clients than if Lackie worked alone. With more minds on a project, they can be more complex and go beyond Lackie's technical skills. Finally, the students gain valuable practical work, data, and programming experience. Outside of Lackie's salary, these deliverables are funded solely by the \href{https://www.carleton.edu/student-employment}{student-employment program} at Carleton College. 

The following language is used in DataSquad student-worker advertisements. It sufficiently introduces the purpose and philosophy of the DataSquad:
\begin{displayquote}
Data and data problems are everywhere. Our aim is to make research and work life easier on campus.  The DataSquad can grapple with simple to thorny data tasks and problems, freeing our clients up for other tasks. Working alone, or in small project-focused groups, DataSquad members will utilize and build on their programming and or data analysis skills to solve data problems or solve problems with data (typically in languages that include Python, R, Java, Javascript, Google Apps Script, and SQL). Additionally, DataSquad members will develop associated supporting materials following the FAIR data principles; project documentation, stakeholder-focused communications, and technical reports.\cite{lackie_22-26_2023}
\end{displayquote}
While the DataSquad had initially focused on issues of "data," our services have expanded to include more expansively data-related projects such as:
\begin{itemize}
    \item  Capture and interpretation of Nintendo Wii\textsuperscript{®} pressure plate data for a cognitive psychology research exploration. 
    \item Recreating an interactive web sankey diagram that links the majors of alumni to their career areas, ultimately linking current students to specific alumni who volunteer to provide career support.
    \item Capturing and re-aligning JSON-formatted closed-caption broadcast news into text files of complete newscasts, resulting in >100,000 text files. 
    \item Creating an interface to more easily allow carleton.edu account holders to identify and preserve files shared with them by accounts of people no longer at the institution.
\end{itemize}

\subsection{Student Roles}

The roles within the DataSquad are organized in a semi-hierarchical way. Responsibilities tend to increase from Technical Writers to Programming Assistants to Lead Data Scientists, with the single Project Management Intern learning to pull it all together. 

Each role can be summarized as the following:
\begin{displayquote}
    \textbf{Technical Writers} will manage and compose technical documents. They are responsible for maintaining the DataSquad blog and website. Writers are often asked to communicate with individual project members to assure that each project has appropriate documentation. 

    \textbf{Assistant Data Scientists} will wrangle, sculpt, develop, and/or visualize a wide range of data sources and formats, often working in small teams. The main goal is to take information from its source and programmatically convert it into formats suitable for the stakeholder/client.
    
    \textbf{Lead Data Scientists} will lead data-related projects. In addition to the responsibilities of Programming Assistants, the Leads will parse projects into achievable steps, assure that documentation is complete, and work with the DataSquad Project Management Intern to archive the project upon completion. 
        
    \textbf{The Project Management Intern} will work primarily with Lackie to help manage the students and projects for the DataSquad. They will be expected to build their skills in team and project organization, communication, and scheduling. Along with Lead Data Scientists, they will parse projects that wrangle, sculpt, develop, and/or visualize a wide range of data sources and formats.  Additionally, Project Management Intern work will include leading subteams of DataSquad members, blogging about projects or internships, and promoting the DataSquad and its activities.
\end{displayquote}
There is additional complexity beyond this order. For example, although Technical Writers may end up writing code for a variety of related projects, they are not specifically charged with programming tasks.  Assistant and Lead Data Scientists expect to be programming most of their time, but often find that they must communicate with their project group and end up writing documentation. In practice, students in all roles get exposure to all key work skills.  The only students who miss out are those who, due to scheduling or temperament, end up working alone most of their time. 

\subsection{Complementary Work Skills}

The four basic roles are designed to complement one another and provide a team-based approach to project work.\footnote{The names of the four roles within the DataSquad model have shifted over time. Specifically, the Assistant Data Scientists have been called Programming Assistants, and Lead Data Scientists have been variously referred to as Data Analysts, Senior Designers, and Team Leaders.} The Technical Writer position was initially designed to give a student of any background a chance to get exposure to data science work. A main purpose of this role is to have the self-identified programmers learn to communicate about their work clearly enough so that a novice Technical Writer can do their job and write it up.  The "novice" may or may not have a programming background, but it is irrelevant; programmers need to be able to describe their logic and often do not realize that they cannot - until they have to. Meanwhile, the idea is that Technical Writers also learn about the programmer's thought process, expand their own technical skills, while gaining valuable writing experience. It has proven challenging to help student programmers appreciate the imperative of clear communications in their own work. The processes of learning to communicate in this way can sometimes cause stress to both the programmer and the Technical Writer beyond productive learning. Consequently, we have occasionally modified expectations; used LinkedIn Learning offerings to better prepare the Technical Writers to see the point of their roles and periodically held "blog-post-writing" events to fill in our stories.  Subsequently, Lackie has employed strategies that reiterate the many ways in which communication skills matter in all post-college work and this has helped to increase appreciation for the important work of documentation. 

Another example can be found in maintaining the \href{https://carletondatasquad.bitbucket.io/}{Carleton DataSquad website}: a 100\% student managed and run project. While some new Technical Writers have chosen to completely re-design and re-write the site documentation, others have broken it – in a variety of ways.  Inheriting someone else's logic turns out to be difficult! The motivation to "make it better" must always be paired with clarity in documenting the process.  This iterative experience of writing on top of other people's documentation leaves a lasting impression, allowing students to appreciate the utility of investing in good documentation. 

The Project Management position was developed with Career Center support and occasional mentoring from certified project managers elsewhere on campus. Beyond the job description responsibilities, students in this role help keep the GitHub\footnote{Because this is a space where novices are getting their feet under them, we keep the Repo private. Largely due to influences from attending USRSE25 Lackie and the Career Center are developing an extracurricular program to support students' learning useful Git skills.} sorted, help Lackie by providing a window into the current student experience, moderating her expectations, and interpreting student behavior. In return, the intern receives peer-supervisory experience, mentoring, and a more realistic view on post-college work life.  

Another feature of the DataSquad model that is designed to safely give DataSquad members memorable work-skills has to do with the different ways students are responsible for some foundational materials of managing the DataSquad itself.  The DataSquad Handbook was originally written by Lackie - but has always been shared as perpetually in draft mode; students are not expected to read it, instead, they are asked to \textit{edit} it. In this way it became more of a social contract and those who did read it have referenced sections in conversation.  There are also some easter eggs in the text to help keep up with who is reading and who has missed the point.  There is much to learn while in college, and few students will have the capacity to absorb everything.  Some students have perceived this participatory design as being disorganized.  These students will, no doubt, \textit{eventually} encounter the learning opportunities afforded by productively engaging in the development of an organizational structure.  Doing so for the first time, while on the job, can be stressful. 

Students are also expected to follow some basic communication protocols: 
\begin{itemize}
    \item Maintain their hours on the staff Google Groups™ Google Calendar™
    \item Respond that they have at least seen a Slack\textsuperscript{®}\footnote{Slack is a registered trademark and service mark of Slack Technologies, Inc.} message by tapping an emoji response (This is apparently harder than it seems.)
    \item Maintain weekly mini-check-in meetings with Lackie or the Project Manager  
    
Throughout these meetings and in all communication opportunities, Lackie works to emphasize the importance of recognizing and following \href{https://www.go-fair.org/fair-principles/}{FAIR Data principles} and \href{https://www.unesco.org/en/open-science}{Open Science} protocols. Most of the examples boil down to helping them see their work in a larger context.  In effect, it is a process of helping each student recognize the transience of being a student and grow into seeing themselves as contributing to larger systems.
\end{itemize}

\section{Assessment: How's it Going?}

To assess whether the DataSquad continues to meet its educational objectives and service goals, we developed two Qualtrics\textsuperscript{®}-based assessment instruments. These instruments were created by Lackie in collaboration with two DataSquad members: Ryan Jiang '27 (Economics and Statistics) and Zaeda Peter '26 (Economics). We employed a mixed-methods analytical strategy combining descriptive statistics, inferential tests, and thematic analysis of qualitative responses. While we recognize the limitations of our small sample sizes and the homogeneity of the alumni sample as recent graduates who all worked under similar circumstances while in college; this is an assessment of the DataSquad model at Carleton College. There is also an unavoidable bias in both of the assessment instrument results because of the common denominator that everyone worked directly with or for Paula Lackie. This very likely will have skewed the responses in a positive direction, even though we tried to mitigate bias with a direct request that they provide their most honest and direct feedback. What we have learned may be generalizable, but our specific data are not.

\subsection{Alumni Assessment Development}

The team compiled a list of 104 DataSquad alumni spanning 2010-2025, verified contact information, and gathered current employment plus postgraduate education data through public LinkedIn profiles. A parallel goal included embedding the outdated information displayed on the DataSquad website and, through the instrument, giving those alumni an option to provide updated information about themselves. 

The alumni assessment instrument was designed to capture alumni perception of the impact of their DataSquad (or its immediate precursor) work-study experience on their post-college work lives. More specifically, the assessment served six primary purposes: (1) capturing feedback on valued program elements, (2) gathering current workforce skill requirements,  (3) documenting career trajectories and outcomes, (4) capturing their insights on which attributes from the DataSquad experience should be emphasized for current students, (5) provide alumni an opportunity to remove or update old information about themselves on our website, and (6) update the list of possible alumni mentors, internship hosts, and other helpful student-alumni connections.

The instrument included both quantitative Likert-scale items and open-ended qualitative questions in three sections:

\begin{enumerate}
    \item \textbf{Skills Development Assessment}: Alumni rated their development across eight domains on a five point scale from "No Improvement" to "Yes, a Lot!"  The domains are detailed in the appendix. The skills development domains are: data wrangling/cleaning, statistical analysis, database design/cloud systems, data visualization/storytelling, coding/software engineering, project management/planning, effective teamwork, and technical documentation. Responses used a 5-point scale 
    \item \textbf{Career Influence Assessment}: For each skill domain, alumni rated its career influence using four categories: "Not Applicable," "Not Influential," "Tangentially Influential," and "Very Influential."
    \item \textbf{Learning Environment Assessment}: Alumni evaluated five dimensions of their experience (feeling comfortable, encouraged, included, supported, and valued) using a 4-point scale: "No," "Not Really," "Somewhat," and "Yes." They also had the opportunity to expand on their answers in an open-text box for each category.
\end{enumerate}

Additional items captured their self-reported DataSquad job roles, open-ended reflections on program benefits and memorable experiences, suggestions for improvement, and permission for public use of responses.  

Upon receipt of the results, four separate readers independently classified the alumni open ended responses. Remarkably, all readers independently came up with highly correlated classifications which were easily harmonized into three large categories: comments that reflected appreciating the Foundational Experiences, opportunity to work with Real Problems, and growth in their Communication skills.  See Table \ref{a:tab:qualitative-theme-descriptions} for a deeper description of these classifications. 

The complete survey instrument is available in supplementary materials.

\subsection{Client Assessment Development}

Lackie also created a list of 39 available clients with a reminder of the projects or services provided by DataSquad members over the past 10 years.  This information was combined into customized invitations to complete the much smaller feedback instrument. Of the 39 clients contacted, 22 completed the assessment, providing a 56\% response rate.  

Clients were asked to rate how highly they would recommend the DataSquad to others (10 point scale),  their satisfaction with the service of the DataSquad across four domains (5-point scale): (1) the quality of work, (2) communication and responsiveness, (3) ease of use, and (4) the speed of delivery.  They had the opportunity to expand on their rating in open text boxes and also to provide suggestions to improve the DataSquad model.

\section{Alumni Feedback}

\subsection{Skills Development}

Reflecting on their time on the DataSquad (or its immediate precursor\footnote{Before the DataSquad was fully formalized, students were hired for the purposes of the DataSquad but not officially designated as members.}), alumni reported substantial skill development across all eight assessed domains (Figure \ref{fig:combined-skill-graph}). Technical Documentation showed the highest average development (M=2.42), with 58\% of alumni reporting maximum improvement. Data Wrangling/Cleaning followed closely (M=2.41), and Teamwork ranked third (M=2.28). 

For the purposes of this assessment, communication-focused skills were identified as; Project Management, Teamwork, and Technical Documentation. All three possessed consistently high development ratings (Ms=2.20-2.42), while classic technical skills displayed more variability (Ms=1.57-2.41). Notably, even the lowest-rated skill, Database Design/Cloud Systems (M=1.57), still averaged above the midpoint on the rating scale, indicating moderately high baseline development across all domains.


\subsection{Career Influence}

\begin{figure*}[h!]
    \centering
    \includegraphics[width=\textwidth]{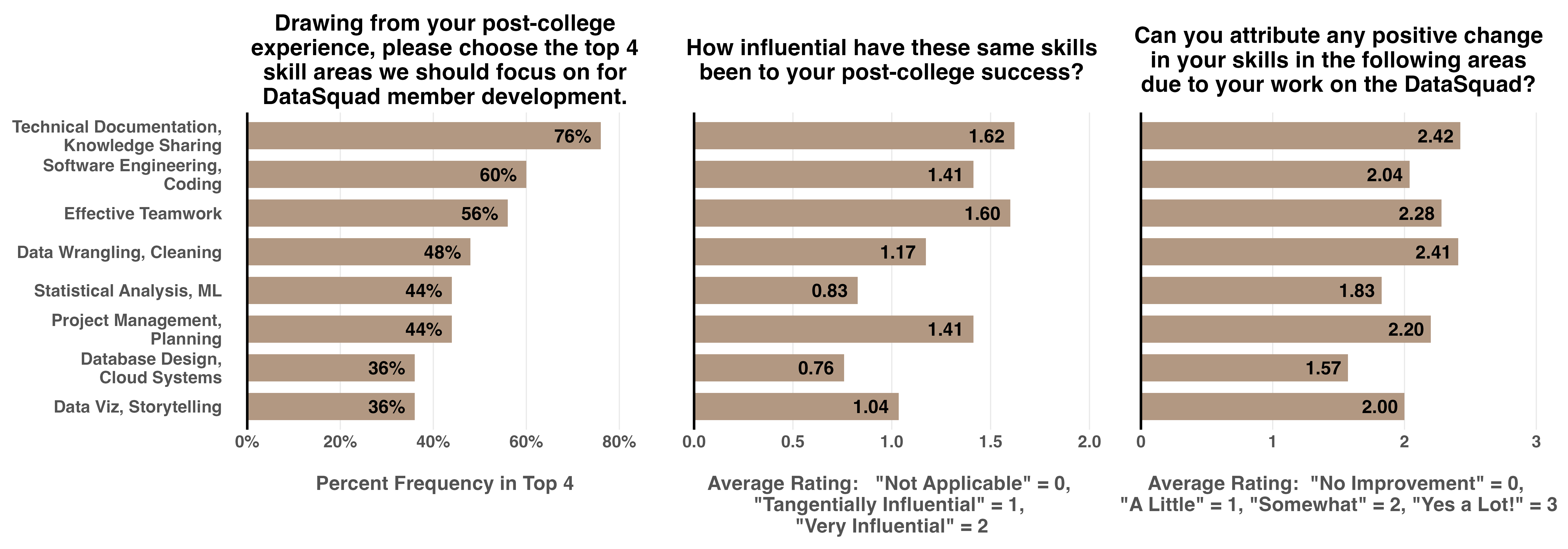}
    \caption{Recommended areas to focus DataSquad efforts, skill influence in post-college success, and skill development while on the DataSquad for alumni. Percent frequency in recommended skill areas is frequency of skill area being placed in top 4. Skill influence and development ratings are averaged across responses.}
    \label{fig:combined-skill-graph}
\end{figure*}

Career influence ratings revealed which skills (from the same list as above) proved most valuable in the professional lives of these alumni (Figure \ref{fig:combined-skill-graph}).  Technical Documentation emerged as most influential (M=1.62), followed by Effective Teamwork (M=1.60) and Coding/Software Engineering (M=1.41).

Overall, communication skills dominated career influence rankings, with three of the top four skills being interpersonal or communication-oriented rather than purely technical. This pattern suggests that while experience on the DataSquad facilitates the development of diverse capabilities, communication skills prove differentiating in professional contexts. Analysis of alumni rankings confirmed this trend: when asked "Drawing from your post-college experience, please choose the top 4 skill areas we should focus on for DataSquad member development." Technical Documentation appeared in 76\% of responses, followed by Software Engineering (60\%) and Teamwork (56\%) (Figure \ref{fig:combined-skill-graph}).  The similarity of these three charts indicates that the alumni felt that through their time on the DataSquad they gained skills they currently value in their careers and want current students to continue to get this experience. 

\subsection{Skill Development and Career Alignment}

\begin{figure*}[t]
    \centering
    \includegraphics[width=\textwidth]{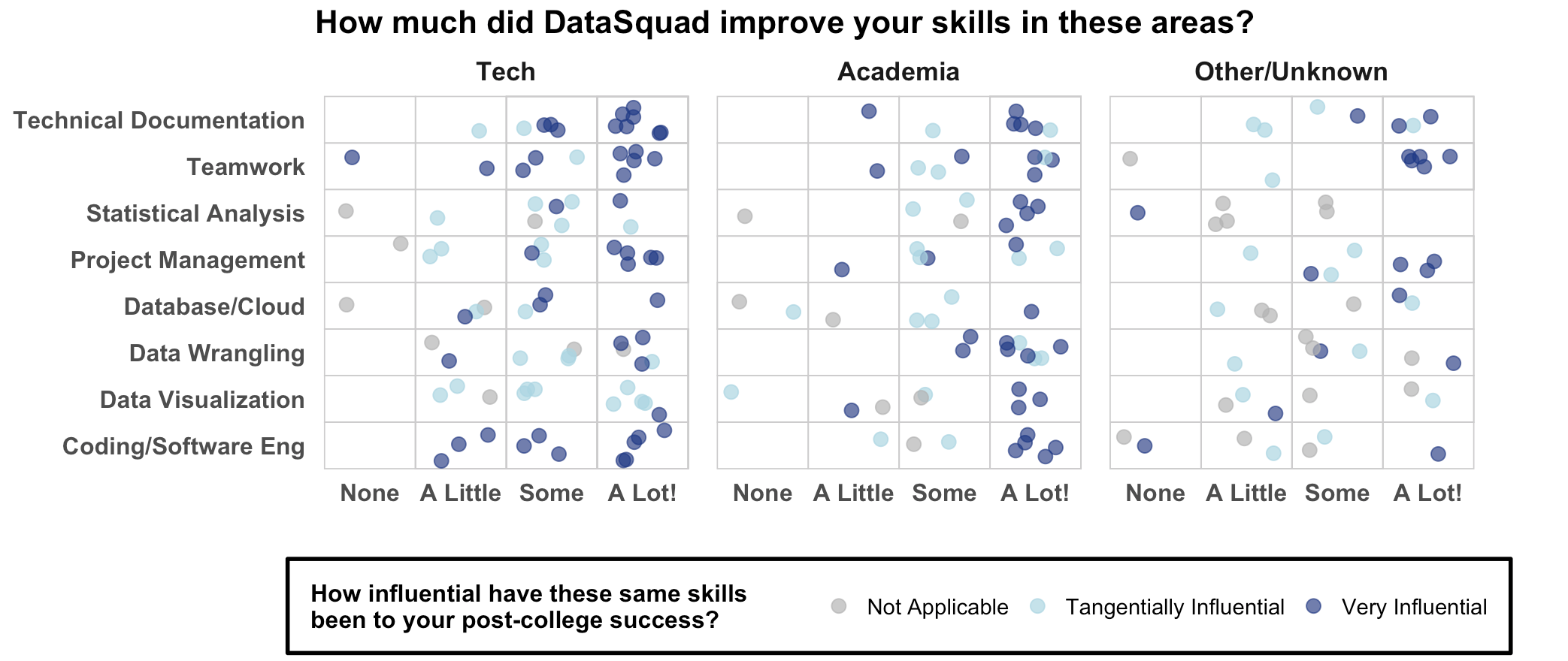}
    \caption{Self-assessed skill development during their time as members of the DataSquad. Responses are broken down by the most recent post-graduate position. Jittered point colors reflect the self-assessed importance of each skill in their post-graduate work.}
    \label{fig:skill-development-v-career-influence-by-career}
\end{figure*}

There is much to glean from Figure \ref{fig:skill-development-v-career-influence-by-career}.  To begin with, we see that DataSquad participants reported meaningful skill development across multiple domains, with the majority indicating at least "some" improvement in key areas. The distribution of responses shows that few participants reported no skill gains, suggesting the program consistently delivers developmental value regardless of career trajectory or the role or roles they held while they were student workers.

Further, the relationship between skill acquisition and post-college influence reveals interesting patterns across career categories.\footnote{These career categories were derived from the public LinkedIn\textsuperscript{®} profiles.} Participants who reported higher levels of skill improvement during their time on the DataSquad ("A Lot!") also tended to rate those same skills as "Very Influential" in their careers; particularly in Tech roles, there is a high concentration of dark dots in the right column of each career category.

Career category differences emerge in both the intensity of skill development and the perceived influence of those skills. While all three career groups show positive skill gains, the concentration of "Very Influential" ratings varies, indicating that DataSquad's impact may be differentially realized depending on career path. The point distributions within each cell in Figure \ref{fig:skill-development-v-career-influence-by-career} reveal the diversity of individual experiences, even among participants with similar overall skill development levels.  

Students who are currently employed in either "tech" or academia present similar patterns of skill development and importance.\footnote{"Tech" includes software engineering, data science, AI/ML roles, and other similar industry positions.} Given that many DataSquad students who go on to pursue graduate education enroll in programs (including computer science, statistics, and economics) with a heavy emphasis on technical ability, this is unsurprising. The students in the "other/unknown" category tend to skew towards business oriented roles in fields like consulting and finance, or find employment working as product managers in tech firms.

\subsection{Qualitative Themes}

The assessment instrument included three open-ended questions exploring skills development beyond coursework, memorable experiences, and suggested improvements. Qualitative analysis of responses to the first two questions (n=17 and n=16 respectively) yielded three main thematic categories, each with multiple subcategories. (See table \ref{a:tab:qualitative-theme-descriptions} for a deeper description of these classifications.)

\begin{figure}[h]
    \centering
    \includegraphics[width=\columnwidth]{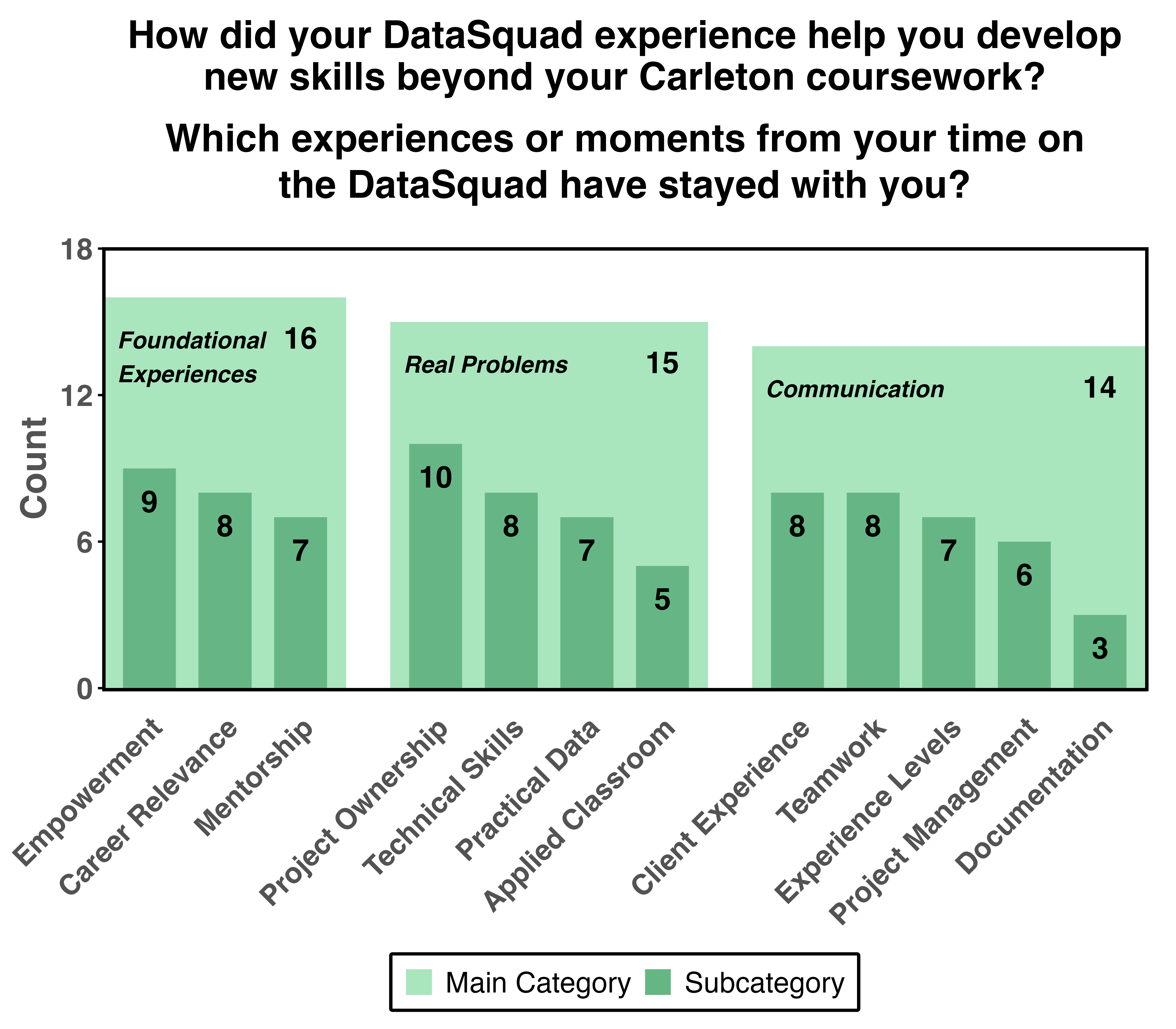}
    \caption{Thematic categories in alumni open-ended responses. Height of light green bars represents prevalence of main categories (count out of 18 total responses). Internal dark green bars show frequency of subcategories. For a single response, neither main categories nor subcategories are mutually exclusive.}
    \label{fig:thematic-categories}
\end{figure}

All three categories showed high prevalence in responses (Figure \ref{fig:thematic-categories}; n=14, n=15, n=16). Among subcategories, the act of leading a project and the confidence gained from working on \textit{real} projects emerged most frequently. Communication improvements—with peers, clients, and across experience levels—appeared consistently across responses, reinforcing quantitative findings about soft skill development.

Drilling into these responses, 14 alumni discussed communication, and there were 32 combined references to specific communication areas (with clients, with teammates, with people across experience level, and with documentation). In these responses, internal and external communication in relation to the DataSquad were both discussed.  The following quotes are examples of these sentiments:

\begin{displayquote}
I learned a lot about professionalism and how to present new ideas to a team. Also helped a lot with developing teamwork abilities. Lastly, I would say it helped me get comfortable asking for help instead of wasting time trying to understand something I have no experience in. (Rajeera Geleta, Computer Science '25)    
\end{displayquote}

Multiple DataSquad alumni  emphasized the importance of learning to communicate across the gap of technical and non-technical backgrounds. Take these examples: Artem Yushko (Computer Science '25) emphasized that "the number one thing was documenting my work and explaining to stakeholders how what I am doing is going to be useful to them"; Elliot Pickens (Mathematics '20) also referenced experience in "communicat[ing] technical knowledge with non-technical clients"; and Quinn Schiller (Computer Science '19) also highlighted his advancements in learning "how to communicate technical work to less technical shareholders."

\section{Client Feedback}



\begin{figure}[h]
    \centering
    \includegraphics[width=\columnwidth]{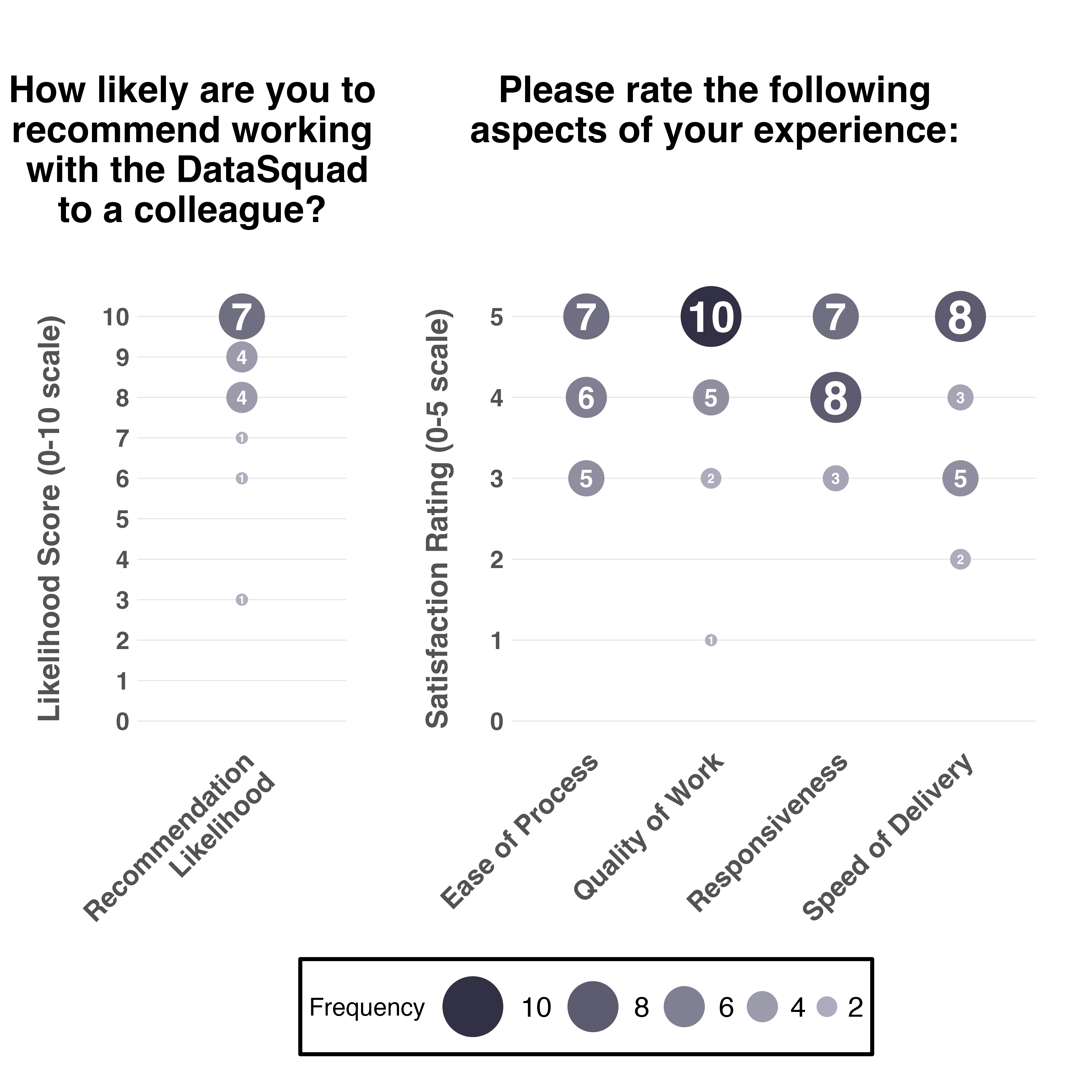}
    \caption{Recommendation likelihood and satisfaction ratings among DataSquad clients. Likelihood to recommend the DataSquad is expressed in a 0-10 scale while satisfaction ratings are between 0 and 5.}
    \label{fig:9}
\end{figure}

The feedback from clients was focused on the delivery of specific service outcomes. DataSquad clients expressed high likelihood at recommending the DataSquad (Fig. \ref{fig:9}; M=8.56). Among clients, satisfaction levels are high as well, with all four subcategories of service evaluation (Quality, Speed of Delivery, Communication, and Ease of Process) having a mean satisfaction of at least 3.9 out of 5. The recommendation question was in a "Net Promoter Score" formatted as 10 stars and embedded in the invitation email, along with a reminder of the project/s and the names of the students who worked on their projects. This format forced a 10-point scale. In this figure, we align the overall score with the 0-5 ranked scales of the 4 domains because the scales are relative to one another. 

When asked to elaborate on their ratings, clients frequently referenced the Speed of Delivery category as an area for improvement (25\% of clients). This is likely due to a combination of Carleton's fast-paced trimester schedule and the 10 hour per week maximum for student-workers, but this is an area that can be improved upon for the DataSquad nonetheless. 

A faculty client provided a nice summary of these issues, while also showing awareness that we're working with students who have fragmented lives. 
\begin{displayquote}
The students on the data squad were busy, so it took a while (speed of delivery) for them to work on my project, and they weren't always super clear about where/what stage they were at. That said, I enjoyed working with them, and there was a moment where [the student] and I were going back and forth and testing the bot that he made to scrape youtube comments that was very fun.  ... I think realistic timelines would help, as well as a recognition that many of us using the DataSquad are humanities people who may not have a lot of familiarity with the tools used by the Squad.
\end{displayquote}

When prompted for suggested improvements to the DataSquad, client suggestions included requesting \enquote{examples of how faculty have used the squad in the past} and proposing a key shift in the mentality of student workers from \enquote{I did what you asked...and should get my A} to \enquote{what is the client really asking for or not asking for, but as a professional I should take the initiative to provide.}

\section{Discussion}

Our analysis of 36 DataSquad alumni over the past 15 years reveals both the sustained effectiveness of this experiential learning model and early indicators of its evolution during a period of rapid technological change. Influential events in this time frame include: Data Science became a career path, the major for Computer Science at Carleton had recently split from the Math department, and Statistics was dubbed as the "Sexy" job for the 2010s.\cite{noauthor_statistics_nodate, press_very_nodate} Naturally, students began to search for \textit{real} applications for their new tech skills.

During the past 15 years, all but one of the students applying to join the DataSquad were looking for the elusive \textit{real }programming experience.\footnote{The one who specifically sought the Technical Writer position was a second trimester Senior Studio Art/Math double major working on rounding out her resume. It was an effective strategy for her.}  The assessment reassuringly confirms the DataSquad model's emphasis on effective communication. In retrospect, alumni report that their practice with a variety of communication challenges while still a college student was not only influential in their post-college lives, but it was ranked higher than any programming/technical skills they gained; they value these communication/soft skills and are clear that they think current students need to gain them as well.

Similar themes are noted from the qualitative assessment of their open ended comments.  These example responses to the prompt "How did your DataSquad experience help you develop new skills beyond your Carleton coursework?" illustrate this alumni view:
\begin{displayquote}
I think working with clients really helped. My current role has a lot of internal client facing work. Having a similar experience definitely helped with my soft skills. Understanding the client's need was important and coping with shifting request happens during real work time.  (Anonymous)\footnote{A 2-year (5 to 10 hours/week) DataSquad veteran who rose to a "Lead" position just before graduating  (Statistics '23) and is now a Data Analyst for a very large corporation.}
\end{displayquote}
\begin{displayquote}
I think the times I spent leading a team of students contributed greatly to my communication and organization skills because they were essential skills that benefited me at work. In addition, talking and working with the client prepared me well and helped me gain the experience I needed. (Joshua Song '23) \footnote{A Computer Science Major and 1-year (10 hours/week) Project Management Intern who went on to be an Implementation Consultant with a software solutions firm.}
\end{displayquote}
\begin{displayquote}
It put me in an environment where I had to start managing not just my own tasks but others (and in a sense, also manage the people), which you can't learn in a classroom. (Veronica Child '18)\footnote{A Computer Science Major and 1-year (10 hours/week) Project Management Intern who went on to be a Software Engineer at Google.}  
\end{displayquote}

The intentional structure of each role can help extend students' focus beyond technical skills and into purposeful transferrable interpersonal skill development. Alumni reported that developing these skills while still a student have a direct payoff in their later work environment.  For instance, 19 out of 25 respondents rated the influence of prior teamwork experience in the workplace as "Very Influential" in their current work-lives. When broken down by current career category, the alumni with tech careers had the highest proportion of "Very Influential" teamwork ratings at 9:1 ("Very Influential" : other).  

Interestingly, while communication was generally satisfactory in quantitative client feedback (Median = 4, 0-5 scale), when given an opportunity to elaborate, clients expressed more critical views. For example, Professor of Spanish Beatriz Parente-Baltran would like more frequent communication about the status of her project and others expressed some frustration over the odd hours of communication, not meeting in person, and generally an interest in the growth potential for more professionalism/follow-through with students. Meanwhile, the alumni highly valued the growth in communication skills that they gained from these very experiences. This divide in opinions reflects the fact that these were the experiences when students learned how to be better at communicating but were not necessarily very good at it at the time. It also reflects the fact that the students, by definition, are so inexperienced as to not be able to judge how long something might take them.  

The client feedback also reinforces the tension that comes out in the IASSIST survey mentioned earlier; can we find a balance between helping our students to be better prepared to tackle \textit{real} problems and also deliver a needed service—one that likely would otherwise languish?  Asking our clients for their feedback has supplied very useful suggestions that will be used to refine the DataSquad Model at Carleton.



\section{Conclusion}


The DataSquad Model provides students an opportunity to balance technical problem-solving with communication (and the other so-called "soft skills") across multiple levels—clients, peers, supervisors. This combination of technical and interpersonal skill-blending is an important phase in the transition from student to professional. Students learn to manage unclear requirements, explain technical decisions and circumstances to non-technical audiences, and negotiate deadlines while maintaining relationships and learning to appreciate what each person has to offer. Although they do not fully realize what it means, they strive to acquire \textit{real} problem solving experience—and that is what they get on the DataSquad.

Not all promising students handle this transition smoothly. Some students excel at coursework but struggle when projects lack clear endpoints or when clients change requirements mid-stream. Others thrive on the ambiguity. Predicting who will adapt and who needs more structured work remains elusive. The program accommodates both paths: students who struggle can step back to more defined roles, while those who adapt quickly, naturally take on greater responsibility.

Despite these varied trajectories, alumni consistently report that the program's combination of technical and collaborative challenges prepared them for professional work. Our survey of 36 alumni spanning 15 years reveals a strong correlation (r=0.78, p=0.024) between skills developed during the program and their subsequent career influence, with Teamwork, Technical Documentation, and Project Management emerging as particularly valuable across career paths.

These findings carry implications for RSE (Research Software Engineering) workforce development. While our alumni pursued varied careers, the skill profile DataSquad cultivates, which combines technical ability with documentation, project management, and collaborative problem-solving aligns closely with competencies that have been identified as essential for research software engineering.\cite{Cohen_2021_four_pill, Goth_2024_found_comp} The model suggests that undergraduate experiential learning, when appropriately supported, can contribute to RSE capacity building earlier in the educational pipeline than traditional graduate-level or on-the-job training.




\section{Acknowledgments}

The 5 months taken to prepare the assessments, work through the Qualtrics processes, analyse the results and write this article have been a tribute to the DataSquad mentality; it has been a thoroughly collaborative experience.  

We thank the 37 DataSquad alumni and 18 clients of those alumni - who generously shared their experiences and insights.  Paula Lackie's vision and leadership over many years created the program this study evaluates; her willingness to subject it to rigorous assessment models the reflective practice we hope to encourage.  

In addition to the authors, we thank the individuals who actively contributed to this article: 
\begin{itemize}
    \item Auiannce Euwing	\textquotesingle26, Assistant Data Scientist: Qualitative analysis and coding to visualize the selected skills Alumni reported as important for Squad member focus
    \item Ryan Jiang \textquotesingle27, Assistant Data Scientist: Qualtrics\textsuperscript{®} development \& metadata collection, data analysis
    \item Zaeda Peter \textquotesingle26, Assistant Data Scientist: Qualtrics\textsuperscript{®} development \& metadata collection
    \item Zoey Fang \textquotesingle28, Technical Writer: Qualitative analysis
    \item William Foote '22 Technical Writer 2021-22 on the UCLA DataSquad: Data analyst for the 2020 IASSIST survey on working with students in data support organizations
    \item     Kristin Partlo, Editor
\end{itemize}

We also thank the attendees of the "People and Networks" session at \href{https://us-rse.org/usrse25/program/}{USRSE25} for their encouraging feedback and enthusiasm for our efforts.\cite{lackie_2025_usrse}

\bibliographystyle{elsarticle-num} 



\bibliography{datasquad_refs}

@misc{lackie_2025_usrse,
  author       = {Lackie, Paula and
                  Pickens, Elliot E.},
  title        = {The DataSquad Experiment:  Some Lessons for
                   Building RSE Capacity
                  },
  month        = {oct},
  year         = {2025},
  publisher    = {Zenodo},
  doi          = {10.5281/zenodo.17503768},
  url          = {https://doi.org/10.5281/zenodo.17503768},
}

@techreport{GeorgetownCEW2024,
  author = {Anthony P. Carnevale, Nicole Smith, Michael C. Quinn},
  title = {Mission Not Accomplished: Unequal Opportunities and Outcomes
for Black and Latinx Engineers},
  year = {2021},
  organization = {Georgetown University},
  url = {https://cew.georgetown.edu/wp-content/uploads/cew-fr-engineering.pdf}
}

@misc{Foote2022-blog,
  author = {Foote, William},
  title = {DataSquad International},
  year = {2022},
  month = {May},
  day = {24},
  url = {https://ucla-datasquad.github.io/#blog},
  note = {Blog post}
}

@article{Kang2022,
  author = {Kang J, Roestel NME and Girouard A},
  title = {Experiential Learning to Teach User Experience in Higher Education in Past 20 Years: A Scoping Review},
  journal = {Frontiers in Computer Science},
  year = {2022},
  doi = {10.3389/fcomp.2022.812907},
  url = {https://www.frontiersin.org/journals/computer-science/articles/10.3389/fcomp.2022.812907/full}
}

@article{Lamprecht2022,
  author = {Lamprecht, Anna-Lena and Martinez-Ortiz, Carlos and Barker, Michelle and Bartholomew, Sadie L. and Barton, Justin and Chue Hong, Neil and Cohen, Jeremy and Druskat, Stephan and Forest, Jeremy and Grad, Jean-Noël and Katz, Daniel S. and Richardson, Robin and Rosca, Robert and Schulte, Douwe and Struck, Alexander and Weinzierl, Marion},
  title = {What Do We (Not) Know About Research Software Engineering?},
  journal = {Journal of Open Research Software},
  year = {2022},
  doi = {10.5334/jors.384},
  url = {https://openresearchsoftware.metajnl.com/articles/10.5334/jors.384}
}

@article{fair-rse,
author = {Anna-Lena Lamprecht and Leyla Garcia and Mateusz Kuzak and Carlos Martinez and Ricardo Arcila and Eva Martin Del Pico and Victoria Dominguez Del Angel and Stephanie van de Sandt and Jon Ison and Paula Andrea Martinez and Peter McQuilton and Alfonso Valencia and Jennifer Harrow and Fotis Psomopoulos and Josep Ll. Gelpi and Neil Chue Hong and Carole Goble and Salvador Capella-Gutierrez},
title ={Towards FAIR principles for research software},

journal = {Data Science},
volume = {3},
number = {1},
pages = {37-59},
year = {2020},
doi = {10.3233/DS-190026},
URL = {  
        https://doi.org/10.3233/DS-190026
},
eprint = { 
        https://doi.org/10.3233/DS-190026
} 
}

@article{Cohen_2021_four_pill,
   title={The Four Pillars of Research Software Engineering},
   volume={38},
   ISSN={1937-4194},
   url={http://dx.doi.org/10.1109/MS.2020.2973362},
   DOI={10.1109/ms.2020.2973362},
   number={1},
   journal={IEEE Software},
   publisher={Institute of Electrical and Electronics Engineers (IEEE)},
   author={Cohen, Jeremy and Katz, Daniel S. and Barker, Michelle and Chue Hong, Neil and Haines, Robert and Jay, Caroline},
   year={2021},
   month=jan, pages={97–105} }

@article{Goth_2024_found_comp,
   title={Foundational Competencies and Responsibilities of a Research Software Engineer},
   volume={13},
   ISSN={2046-1402},
   url={http://dx.doi.org/10.12688/f1000research.157778.1},
   DOI={10.12688/f1000research.157778.1},
   journal={F1000Research},
   publisher={F1000 Research Ltd},
   author={Goth, Florian and Alves, Renato and Braun, Matthias and Castro, Leyla Jael and Chourdakis, Gerasimos and Christ, Simon and Cohen, Jeremy and Druskat, Stephan and Erxleben, Fredo and Grad, Jean-Noël and Hagdorn, Magnus and Hodges, Toby and Juckeland, Guido and Kempf, Dominic and Lamprecht, Anna-Lena and Linxweiler, Jan and Löffler, Frank and Martone, Michele and Schwarzmeier, Moritz and Seibold, Heidi and Thiele, Jan Philipp and von Waldow, Harald and Wittke, Samantha},
   year={2024},
   month=nov, pages={1429} }

@article{wiltshire_using_2025,
	title = {Using the {Datasquad} model to join practical "people skills" with data science education},
	copyright = {Copyright (c) 2025 ISI/IASE},
	issn = {3051-0449},
	url = {https://iase-pub.org/conference_proceedings/IASECP/article/view/117},
	doi = {10.52041/iase24.603},
	language = {en},
	urldate = {2025-09-28},
	journal = {IASE Conference Proceedings Series},
	author = {Wiltshire, Deborah and Lackie, Paula},
	month = feb,
	year = {2025},
}

@article{verma_research_2023,
	title = {Research {Support} {Services}: {An} {Analysis} of top {Science} and {Technology} {Institutions}},
	shorttitle = {Research {Support} {Services}},
	url = {https://digitalcommons.unl.edu/libphilprac/7602},
	journal = {Library Philosophy and Practice (e-journal)},
	author = {Verma, Vijay Kumar and Charu},
	month = apr,
	year = {2023},
}

@incollection{wiltshire_meeting_2025,
	address = {Chicago, Illinois},
	title = {Meeting the {Challenges} of {Data} {Support} {Services} in {Academic} {Libraries}: {Advocating} for an {International} {DataSquad} {Model}},
	isbn = {979-8-89255-616-3},
	language = {en},
	booktitle = {Data {Culture} in {Academic} {Libraries}: {A} {Practical} {Guide} to {Building} {Communities}, {Partnerships}, and {Collaborations}},
	publisher = {Association of College and Research Libraries},
	author = {Wiltshire, Deborah and Lackie, Paula and Dennis, Tim and Parke, Elizabeth},
	year = {2025},
	pages = {209--222},
}

@article{lackie_22-26_2023,
	title = {22-26 {DataSquad} {Job} descriptions-{ITS}},
	urldate = {2025-09-20},
	journal = {Unpublished},
	author = {Lackie, Paula},
	year = {2023},
}

@book{kellam_databrarianship_2016,
	address = {Chicago},
	title = {Databrarianship: {The} {Academic} {Data} {Librarian} in {Theory} and {Practice}},
	url = {https://alastore.ala.org/content/databrarianship-academic-data-librarian-theory-and-practice},
	publisher = {Association of College and Research Libraries},
	author = {Kellam, Lynda and Thompson, Kristi},
	year = {2016},
}

@misc{noauthor_statistics_nodate,
	title = {Statistics {Is} {The} {Sexy} {In} {Science} {\textbar} {Labcoat} {Life} {\textbar} {Learn} {Science} at {Scitable}},
	url = {https://www.nature.com/scitable/blog/labcoat-life/statistics_is_the_sexy_in/},
	language = {en},
	author={Khalil A. Cassimally},
	year={2012},
	urldate = {2025-11-21},
	note = {Cg\_cat: Labcoat Life
Cg\_topic: Statistics Is The Sexy In Science},
}

@misc{press_very_nodate,
	title = {A {Very} {Short} {History} {Of} {Data} {Science}},
	url = {https://www.forbes.com/sites/gilpress/2013/05/28/a-very-short-history-of-data-science/},
	language = {en},
	urldate = {2025-11-21},
	journal = {Forbes},
	author = {Press, Gil},
	year={2013},
	note = {Section: Tech},
}



\newpage
\appendix
\section{Supplementary Results}

\subsection{Learning Environment Ratings}

\begin{figure}[ht!]
    \centering
    \includegraphics[width=\columnwidth]{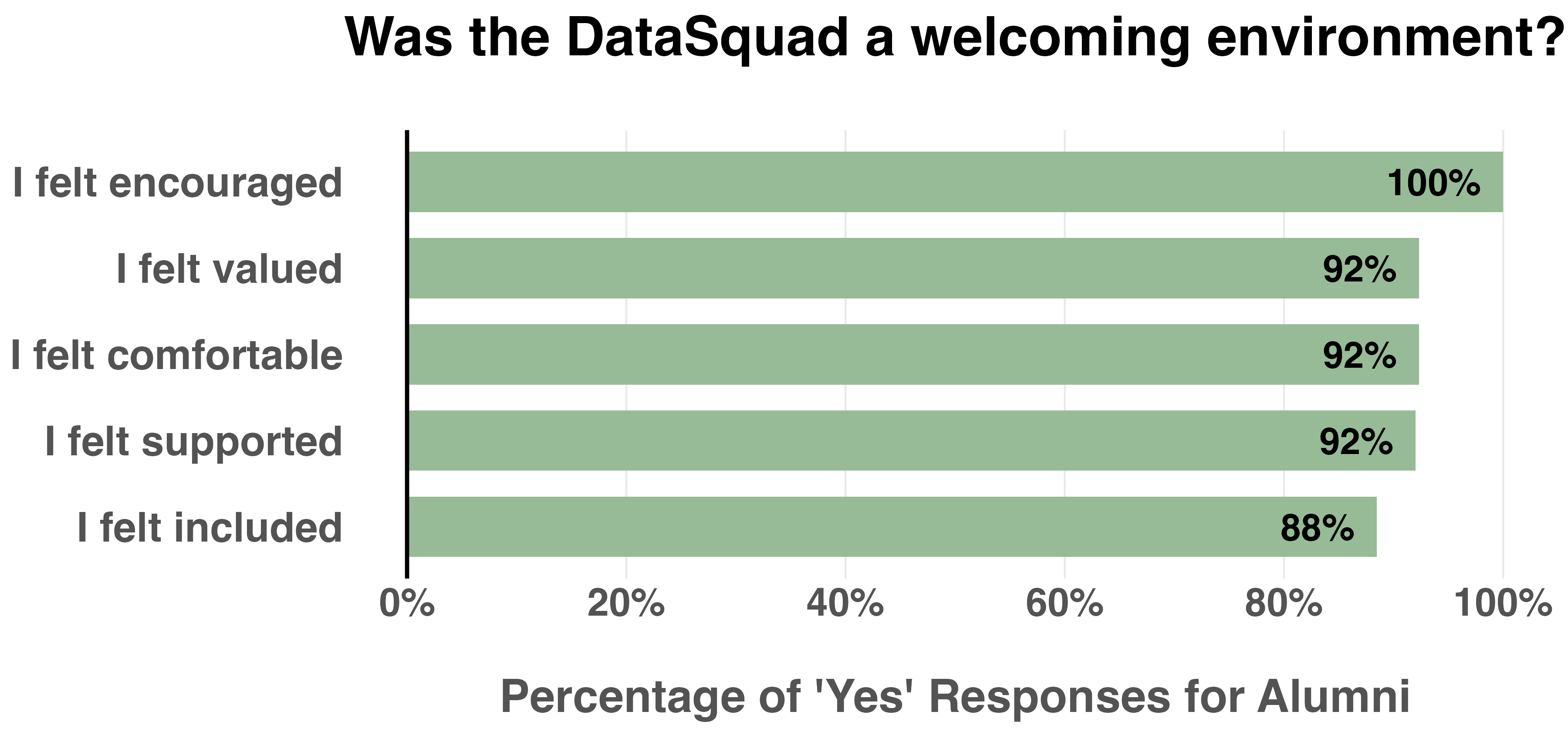}
    \caption{Alumni assessment of the DataSquad environment.}
    \label{fig:environment-assessment}
\end{figure}

The DataSquad learning environment received exceptionally positive ratings across all assessed dimensions (Figure \ref{fig:environment-assessment}). All five dimensions met or exceeded 88\% \enquote{Yes} responses. Although the high ratings may reflect some level of response bias, the 100\% encouragement rate is a positive sign, given that alumni who reported less positive feelings in the other categories agreed on the encouraging environment. It also suggests that there is some work to be done to improve the DataSquad experience for students who responded less favorably.





\subsection{Role Distribution and Multiple Pathways}

\begin{figure}[h!]
    \centering
    \includegraphics[width=\columnwidth]{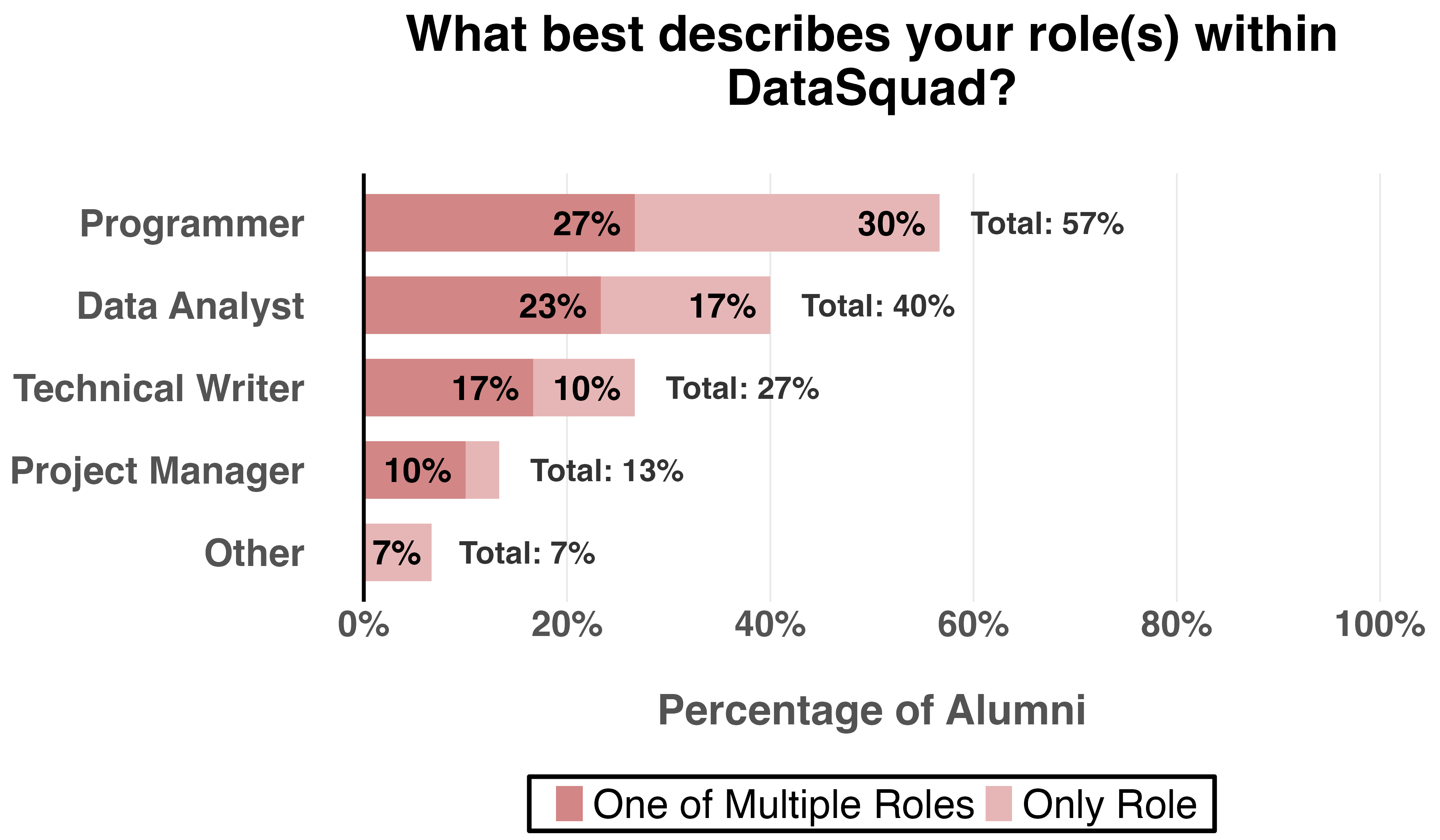}
    \caption{Distribution of roles held by alumni. If multiple roles were held, both are counted and distinguished.}
    \label{fig:A2}
\end{figure}

Many students experienced multiple roles during their tenure (Figure \ref{fig:A2}). Among survey respondents, 57\% served as Programmers/Data Scientists, 40\% as Data Analysts, 27\% as Technical Writers, and 13\% as Project Management Interns. Importantly, approximately half of alumni held two or more roles during their time on DataSquad, gaining breadth across the program's offerings.

\section{Definitions}

Table \ref{a:table:skill-descriptions} shows the "Skill Area" descriptions we provided during the Alumni Assessment instrument.

\begin{table*}[htbp]
\centering
\renewcommand{\arraystretch}{1.2}
\begin{tabular}{|>{\raggedright\arraybackslash}p{4cm}|>{\raggedright\arraybackslash}p{10cm}|}
\hline
\rowcolor{gray!10}
\textbf{Skill Area} & \textbf{Description} \\
\hline
Data Wrangling / Cleaning& \small Scraping or capturing data, setting up a GitHub repository, consulting on data programming languages and strategies. \\
\hline
Statistical Analysis & \small Collecting, exploring and presenting large amounts of data to discover underlying patterns and trends \\
\hline
Database Design, Cloud Systems & \small Designing a safe place to capture your data (in SQL or other), working with data capture or management tools like Qualtrics or Google Forms \\
\hline
Coding, Software Engineering & \small Using programming languages, such as Python, R, etc., and utilizing file management tools like Git\\
\hline
Project Management, Planning & \small Organizing tasks, managing time, and coordinating resources to achieve goals \\
\hline
Effective Teamwork & \small Collaborating well with others, supporting teammates, and achieving shared objectives \\
\hline
Technical Documentation, Knowledge Transfer & \small Creating clear guides, instructions, or documentation so others can understand and use tools, processes, or information; effectively sharing knowledge with teammates \\
\hline
\end{tabular}
\caption{Skill descriptions and definitions.}
\label{a:table:skill-descriptions}
\end{table*}

Table \ref{a:tab:qualitative-theme-descriptions} contains the category descriptions for each of the qualitative categories shown in Figure \ref{fig:thematic-categories}.

\begin{table*}[htbp]
\centering
\renewcommand{\arraystretch}{1.2}
\begin{tabular}{|>{\raggedright\arraybackslash}p{4cm}|>{\raggedright\arraybackslash}p{10cm}|}
\hline
\rowcolor{gray!10}
\textbf{Category Name} & \textbf{Description} \\
\hline
Empowerment & \small Feelings of accomplishment upon completion or participation in DataSquad projects\\
\hline
Career Relevance & \small Development of skills that have a tangible impact in the workplace \\
\hline
Mentorship & \small Guidance from Paula in navigating Carleton and post-college paths \\
\hline
Project Ownership & \small The act of owning and meaningfully contributing to a project\\
\hline
Technical Skills & \small Raw technical skill development, commonly tied to certain programming languages \\
\hline
Practical Data & \small Real life experiences with messy data \\
\hline
Applied Classroom & \small Complementary, real-world experience that pairs with academic studies \\
\hline
Client Experience & \small Interacting with the client, real world exposure at delivering products and services \\
\hline
Teamwork & \small Internal communication within the DataSquad \\
\hline
Experience Levels & \small Communicating with clients and squad members of different tech backgrounds \\
\hline
Project Management & \small Supervising and coordinating a project's deliverables from start to finish \\
\hline
Documentation & \small Collecting and recording data, documents, and other project information throughout the lifecycle of a project \\
\hline
\end{tabular}
\caption{Alumni Qualitative Theme Descriptions}
\label{a:tab:qualitative-theme-descriptions}
\end{table*}

\end{document}